*Data science skills for referees: I Biological X-ray crystallography*

John R Helliwell

School of Chemistry, University of Manchester, Manchester M13 9PL, UK

*Synopsis*

There is now a growing wish by referees to judge the underpinning data for a submitted article. It is timely therefore to provide a summary of the data evaluation checks requiring to be done. As these will vary from field to field this article focuses on the needs of biological X-ray crystallography articles, which is the predominantly used method leading to depositions in the PDB.

**Abstract**

Since there is now a growing wish by referees to judge the underpinning data for a submitted article it is timely to provide a summary of the data evaluation checks required to be done by a referee. As these checks will vary from field to field this article focuses on the needs of biological X-ray crystallography articles, which is the predominantly used method leading to depositions in the PDB. The expected referee checks of data underpinning an article are described with examples. These checks necessarily include that a referee checks the PDB validation report for each crystal structure accompanying the article submission; this check whilst necessary is not sufficient for a complete evaluation. A referee would be expected to undertake one cycle of model refinement of the authors' biological macromolecule coordinates against the authors' processed diffraction data and look at the various validation checks of the model and Fo-Fc electron density maps in e.g. Phenix_refine and in COOT. If the referee deems necessary the diffraction data images should be reprocessed (e.g. to a different diffraction resolution than the authors' submission). This can be requested to be done by the authors or if the referee prefers can be undertaken directly by the referee themselves. A referee wishing to do these data checks may wish to receive a certificate that they have command of these data science skills. The organisation of such voluntary certification training can e.g. be via those crystallography associations duly recognised by the IUCr to issue such certificates.

1. **Introduction**

The level of rigour for refereeing the data underpinning a submitted article in all areas of science is receiving increasing attention. The degrees of such rigour have been classified into four types (see e.g. http://www.springernature.com/gp/group/data-policy/policy-types). The most rigourous is defined as Type 4 for which "a journal would require that all datasets on which the conclusions of a paper rely be available to reviewers and readers. Authors must deposit their datasets in publicly available repositories prior to peer review, or include them as supplementary information files with their manuscript. It is a condition of publication that authors deposit their data in an appropriate repository, and agree to make the data publicly available without restriction, unless reasonable controls on data access are needed to protect human privacy or biosafety." Similarly the International Council for Science points out that "Publishers have a responsibility to make data available to reviewers during the review process." (http://www.icsu.org/science-international/accord/open-data-in-a-big-data-world-long ).

Currently biological crystallography is lagging behind chemical crystallography, which is achieving the Type 4 level of rigour, at least for IUCr journals. It is excellent that the PDB has introduced a Validation report and a Validation service. This is precisely what I pushed for along with IUCr Chester during my time as Acta Cryst Editor in Chief (1996-2005). PDB has responded. This is one half of what is needed in my view. The other half of the required procedure, clearly envisaged within the Type 4 concept, is that referees of biological crystallography articles should routinely also have access to the diffraction data and coordinates as lodged with PDB, as well as the PDB's Validation report.

In judging the data underpinning an article the referee will be assessing the procedure that authors will have followed from raw data to processed structure factors to finalised molecular model coordinates. There are numerous steps and various software programs that can be used. A sensible view will need to be taken of the author's steps, and which may not be the preferred steps that the referee would have taken. There will then be a variance that can be allowed. *Outside that variance however errors can be determined and which should be clearly described in the referee's report to the editor.* The descriptions below aim to provide advice to referees of how to evaluate such data involving PDB depositions.

An article will no doubt have particular results to report and the underlying data will be the underpinning evidence. But once in the PDB a wider use of the data might be possible. It is reasonable for a referee to point out deficiencies in either the processed diffraction data or derived molecular model to an editor so as to gain utility beyond the results in an article. Such an approach has worked well in chemical crystallography where that philosophy has been applied.

Let's take an overview now of the PDB Validation report contents and the calculational checks that a referee will make of their own.

Editors of even non-specialist journals now realize that requiring as a minimum the PDB validation reports (Read et al 2010) might prevent the most obvious transgressions of biological macromolecule model plausibility. Thus, for example, Fink (2016) states "The *Journal of Immunology* now requires that the PDB Summary Validation Report …be included with submission of the manuscript so that it is available to editors and reviewers during the review process."

Read et al 2010 have a section especially tailored to advising referees of submitted PDB files and quoting directly, transposing their future tense to present tense syntax:- *"The first page (of the PDB validation report) gives an overall summary, with key percentile scores for global quality on both all-PDB and resolution-relative scales. The first page also gives any "concerns" or "unusual features" present in the structure or data and gives per-chain quality indicators, including mean B-factor, overall RSR-Z, and overall RMS-Z for bonds, angles, and planes. Subsequent pages provide detailed information on residue-based quality indicators, allowing the referee to assess the level of confidence for specific residues discussed in the manuscript."*

Whilst excellent in providing a uniformity of what is assessed unfortunately, in many situations the PDB Validation report is insufficient to pinpoint the validity of an article's claims and models based on specific electron density interpretations. From a recent informal polling of colleagues a wide variety of journal editors forward to manuscript authors any referee's requests for structure factors and model coordinates to allow full validation of the claims. These requests by referees for access to the underpinning data for a submitted article are now growing in number, whilst not compulsory this is perceived by many referees as 'good practice' and which I obviously agree with. It is timely then to provide a discussion document of the data evaluation checks required to be done by a referee. This will also guide the crystallography associations in their plans for future continual professional development (CPD) of their members and certifying their data science skills.

2. *Methods and Examples*

*2.1 Refereeing of data and atomic coordinates in practice*

There can only be a realistic burden of work on referees in practice. Authors must remain in charge of their papers. But structural biology as a field **cannot expect a privilege of exemption** from refereeing of the diffraction data and derived model coordinates underpinning a publication.

Fortunately with the current state of computer laptops and PCs a single cycle of re refinement of a model using eg Phenix Refine (Afonine et al 2012) yields a readily accessible suite of validation checks of that model and underpinning electron density maps, displayed eg in COOT molecular graphics (Emsley et al 2010). So, reasonably easy interrogation in specific detail is therefore possible by a referee.

Where there are concerns from a referee to do with significant difference electron density revealed via their own calculations these would be flagged explicitly by the referee for the attention of the handling Editor for that article, and which could state, as in Acta Cryst C and checkcif:- *'Alert level A: Must be attended to by the authors before acceptance of the article could be considered'*.

*2.2 Examples of validation information available from Phenix Refine and displayed below as a mock up of a referee's report on an article's underpinning diffraction data and coordinates*

A referee would need to declare that they have read the PDB validation report(s). An example comment to this report might for example be: "This showed several amino acids with RSRZ values flagged as high; these residues need scrutiny by the authors and if no correction proves possible a comment should be included in a Supplementary file to the article."

The rest of a referee's report, as illustrated below, would be based on their own direct calculation checks and would be in addition to their scrutiny of the PDB Validation report described above.

A referee would need to declare that they undertook a round of model refinement using e.g. Phenix Refine (Afonine et al 2012). The Fo-Fc difference electron density map peaks list displayed in COOT if it includes some large peaks that need attention then a comment can be requested be made in a Supplementary file to that article by the authors, that is, if no reasonably certain chemical assignment is possible.

The example shown in Figures 1, 2, 3 and 4 is 'anonymised' and used for illustration purposes only.

Where there are a quite a number of bound solvent water molecules in the PDB deposited model for which the 2Fo-Fc electron density is below 1.2 rms these should be reviewed as to whether they should really be included in the bound water structure in the PDB file. A further check is the COOT validation of bound waters; this is also informative about questionable waters.

In some cases raw diffraction images may need to be requested via the editor to the authors if the referee deems necessary. This may be necessary for example if the 'Table 1 Model refinement summary' of the article states that truncation of the diffraction data resolution at 'too high' a value of <I/sigI> and / or $CC_{½}$ has been made by the authors. The merits of harnessing the diffraction beyond the 'conventional' resolution limit in the case of biological macromolecule model refinement have been described by Einspahr and Weiss (2011) and by Diederichs and Karplus (2013).

### 2.3 Assessing the precision of displayed distances for non-covalent interactions

The figures of an article that depict non-covalent interactions need to be vetted to a proper precision of any displayed distances. A website tool for achieving this vetting is now available and specific types of examples are provided (Kumar et al 2015).

### 2.4 A simple viewing by a referee of the derived coordinates and processed diffraction data files

For a referee to simply open the coordinates file and scroll through and read it can reveal unexpected things. One such is the peculiar case of 1lkr where individual amino acid side chain atoms had erroneously variable occupancies, which is obviously physically meaningless. This had not been noticed by anyone in the writing of the paper by the author nor the referees or by the editor or the PDB annotator handling it. For a full description of the case of 1lkr as an example see Helliwell and Tanley (2016).

Another relatively simple data check is to open the processed diffraction data and examine the 'zero layer diffraction zones'. This is straightforward to do using the software suite CCP4 for example when opening an 'mtz' file (Winn et al 2011). This reveals gaps in the data that might exist due for example to physical obscuration of the apparatus of the diffracting solid angle or due to ice rings.

### 3. Discussion

The Methods and Examples section above documented various areas for data and calculation checks by referees. There are some aspects however where, even amongst highly experienced crystallographers, there is not necessarily a consensus. This can also lead to variations between journal editors as well as authors. These would include:-

* the number of assigned bound waters

* the number of assigned alternate (i.e. split) occupancy amino acid side chains

* the number and type of assigned anions and cations. [The particular category of decision making involving the low atomic number ions is often exacerbated by insufficient experimental data but which is set to improve by new user beamline provision involving utilising long X-ray wavelengths with leadership by Diamond Light Source of the new beamline I23 led by Dr Armin Wagner http://www.diamond.ac.uk/Beamlines/Mx/I23.html . ]

* Processed diffraction data (poor) completeness affects the precision of a refined molecule. Beyond this plain statement there is no community consensus however on what value is reasonable. However it is fair to report that a high resolution shell dropping below 50% in its completeness should at least be commented upon in an article and it should in any case not be allowed to form the basis of a claim for the quoted resolution limit in the title or abstract.

* For electron density map contouring there is no community consensus on the sigma level above which the map is contoured. Typically in publications, indicative therefore of a consensus, Fo-Fc maps are shown contoured from 5 sigma (this being the COOT default value) but from 3 sigma contour level is also common. Anomalous difference Fourier maps are displayed contoured from 3 sigma typically. 2Fo-Fc maps are shown contoured from 1.2 rms. Beyond this there are other practices however. The 2Fo-Fc map can be shown contoured at much lower levels, such as down to 0.8 rms, and where continuity of the map is taken to support an interpretation of a particular molecule or molecular fragment being present. This is questionable practice and a referee would be reasonable to ask that a second crystal be analysed and the same portion of an electron density map be scrutinised. If a basically identical result is obtained then statistically speaking the certainty of the initial interpretation is more confident. Further confidence in a difference map can be obtained by the referee if the same diffraction data is used within two different software packages.

4. **Guide to the referee's recommendation options on the data sets underpinning an article**

These are really not different from the recommendations that a referee makes to an editor regarding a submitted article namely 'Accept', 'Minor revisions', 'Major revisions' or 'Reject' and will most likely be fairly obvious to the referee in any given case, or decided upon in consultation with the editor.

Overall the referee is not the author and whilst sufficient excursions into the data as described above are made by the referee the actual work of revisions of the data sets, atomic coordinates or structure factors or even raw data remeasurement, is obviously the task of the authors. A list of recommendations to the editor for revisions to be made by the authors would be given.

## 5. Concluding remarks

The expected referee checks of data underpinning an article have been described with examples. These checks necessarily include that a referee checks the PDB validation report for each structure accompanying the article submission; ***this check whilst necessary is not sufficient for a complete evaluation***. A referee would then be expected to undertake one cycle of model refinement of the authors' biological macromolecule coordinates against the authors' processed diffraction data and look at the various validation checks of the model and Fo-Fc electron density maps in eg Phenix_refine and in COOT. If the referee deems necessary (i.e. in the referee's judgement based on the above summarised work) the diffraction data images should be reprocessed (e.g. to a different diffraction resolution than the authors' submission). This can be requested be done by the authors or if the referee prefers can be undertaken directly by the referee themselves.

A referee wishing to do these data checks may wish to receive a certificate that they have command of these data science skills. The organisation of such voluntary certification training can be via the crystallography associations, approved by the IUCr, and their authorised Continual Professional Development 'CPD' committees.

**Acknowledgements**

JRH, as Editor in Chief of IUCr Journals 1996 to 2005 saw in detail the exemplary checks undertaken by Acta Crystallographica Section C Coeditors, which included Dr Madeleine Helliwell, using the checkcif report as well as the chance to scrutinise with their own calculations the underpinning processed structure factors and derived atomic coordinates of each submitted article. Thus JRH is very grateful to all colleagues involved with Acta Crystallographica Section C at the time notably Professor Syd Hall and also Dr Madeleine Helliwell for many discussions.

Figures

Figure 1 An example of a Fo-Fc peaks list displayed in COOT clearly showing the need for further attention by authors.

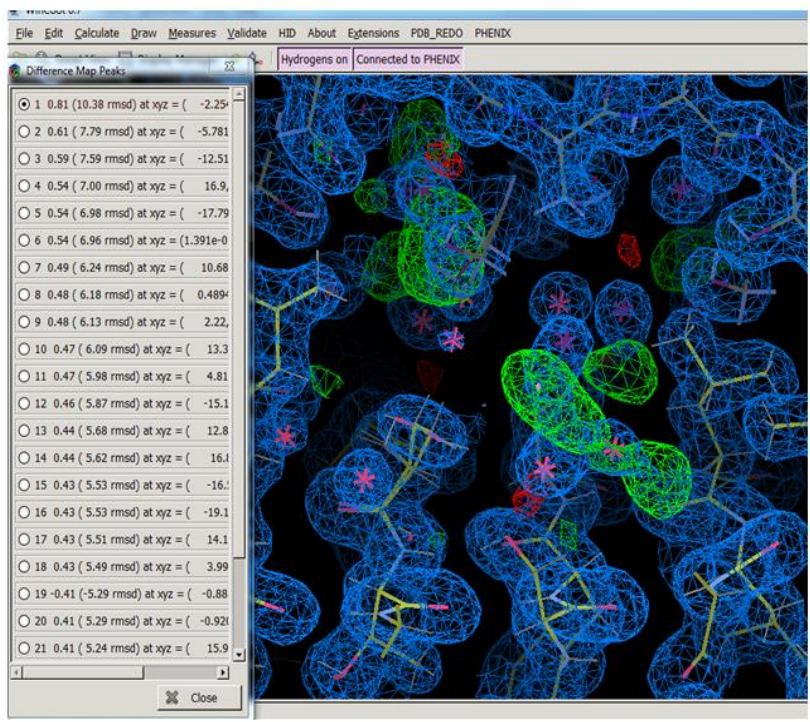

Figure 2 An example of a Phenix_refine diagnostics plot showing an unusually high average B factor clearly showing the need for further attention and/or comment in their article by the authors.

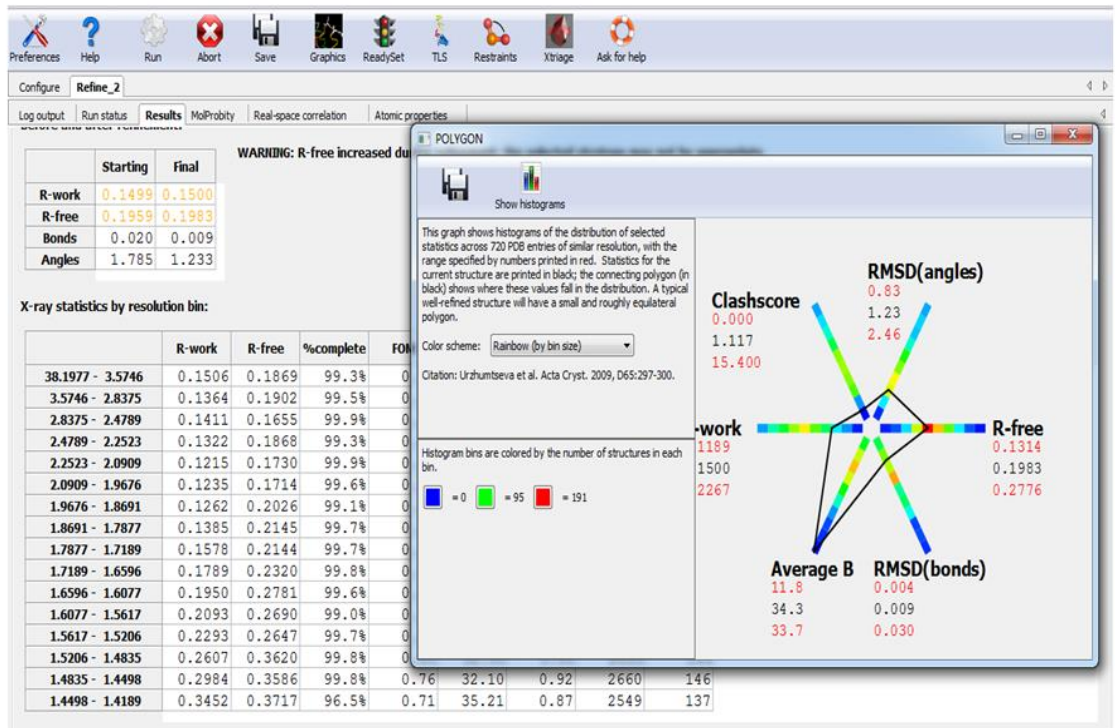

Figure 3 Molprobity diagnostics also available within Phenix Refine indicate clashes for several water molecules and which need correcting, clearly showing the need for further attention by authors.

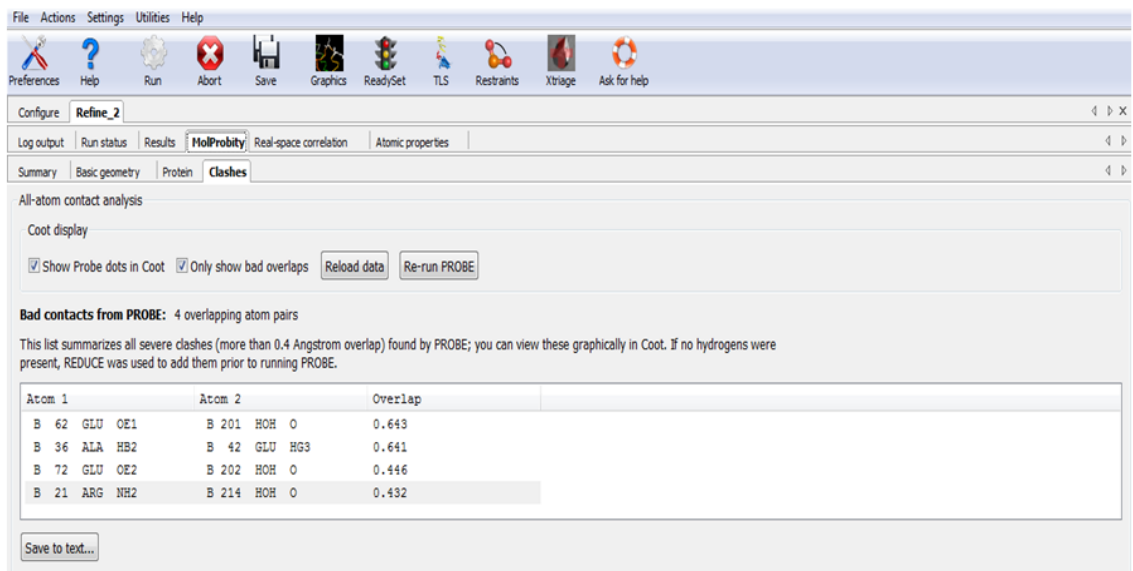

Figure 4 Phenix_refine shows that there are several amino acid residues with very high B factors that need authors' further attention or comment made in a Supplementary file to the article if no improvement is possible by the authors.

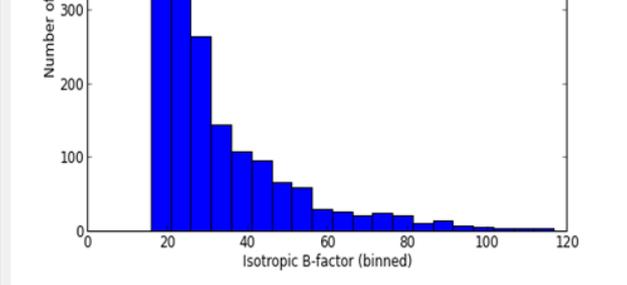

**Suspiciously high B-factors**

The table below lists all isotropic ADPs with values greater than four standard deviations above the mean value for this structure. Although a high B-factor is not necessarily wrong, it may be worth double-checking the atomic positions, occupancies, or (rarely) element types.

| Atom ID | Isotropic B-factor | Occupancy |
|---|---|---|
| NH1 ARG A  21 | 116.73 | 1.00 |
| NH2 ARG A 104 | 114.59 | 1.00 |
| NH1 ARG A 104 | 113.05 | 1.00 |
| CZ  ARG A 104 | 111.40 | 1.00 |
| NH1 ARG B 104 | 110.11 | 1.00 |
| NH2 ARG B 104 | 107.10 | 1.00 |
| NH2 ARG B 103 | 105.81 | 1.00 |
| CZ  ARG B 104 | 105.00 | 1.00 |